\documentclass{article}

\usepackage[preprint]{neurips_2026}

\usepackage[utf8]{inputenc} 
\usepackage[T1]{fontenc} 
\usepackage{hyperref} 
\usepackage{url} 
\usepackage{booktabs} 
\usepackage{amsfonts} 
\usepackage{nicefrac} 
\usepackage{microtype} 
\usepackage{xcolor} 

\hypersetup{hidelinks,breaklinks=true}
\usepackage{tabularx}
\usepackage{array}
\usepackage{xurl}
\usepackage{enumitem}
\usepackage{caption}

\usepackage{tikz}
\usetikzlibrary{arrows.meta,positioning,calc,fit,backgrounds}
\definecolor{cdink}{gray}{0.15}
\definecolor{cdline}{gray}{0.45}
\definecolor{cdfill}{gray}{0.94}
\definecolor{cdacc}{rgb}{0.05,0.32,0.40} 
\definecolor{cdaccfill}{rgb}{0.90,0.945,0.95}
\tikzset{
 cdbox/.style ={draw=cdline, fill=cdfill, rounded corners=1.5pt, inner sep=3pt,
 align=center, font=\scriptsize, text=cdink},
 cdmodel/.style ={draw=cdink, line width=0.6pt, fill=white, rounded corners=2pt,
 inner sep=4pt, align=center, font=\small},
 cdacq/.style ={draw=cdacc, fill=cdaccfill, rounded corners=1.5pt, inner sep=3pt,
 align=center, font=\scriptsize, text=cdink},
 cdarr/.style ={-{Stealth[length=4pt,width=3pt]}, draw=cdline, line width=0.5pt},
 cdarr5/.style ={-{Stealth[length=4pt,width=3pt]}, draw=cdacc, line width=0.7pt},
 cdtag/.style ={circle, draw=cdline, fill=white, inner sep=0.7pt, font=\scriptsize\bfseries,
 text=cdink},
 cdtag5/.style ={circle, draw=cdacc, fill=white, inner sep=0.7pt, font=\scriptsize\bfseries,
 text=cdacc},
 cdband/.style ={font=\scriptsize\itshape, text=cdline},
 cdstep/.style ={cdbox, text width=4.35cm,
 font={\scriptsize\hyphenpenalty=10000\exhyphenpenalty=10000}},
}

\setlist{topsep=2pt,itemsep=2pt,parsep=0pt,leftmargin=1.4em}

\newcommand{\code}[1]{\texttt{\small #1}}

\newcommand{\auditseed}{seed = 20260812}

\newcommand{\rDocs}{29} 
\newcommand{\rOrgs}{26} 
\newcommand{\rKw}{0.46} 
\newcommand{\rKwLo}{0.37} 
\newcommand{\rKwHi}{0.54} 
\newcommand{\rKwMin}{0.00} 
\newcommand{\rKwMinVar}{F3 t2 and t4, where almost every code is \code{0}} 
\newcommand{\rKwMed}{0.21} 
\newcommand{\rKwMax}{0.35} 
\newcommand{\rKwMaxVar}{F4 Regeneration}
\newcommand{\rRaw}{65} 
\newcommand{\rACone}{0.56} 
\newcommand{\rIntraKw}{0.84} 
\newcommand{\rRateFone}{39} 
\newcommand{\rRateFtwo}{13} 
\newcommand{\rRateFfour}{11} 
\newcommand{\rRateTone}{5} 
\newcommand{\rRateTfive}{5} 
\newcommand{\rAdjud}{98} 

\newif\ifcodersnamed
\codersnamedfalse
\newcommand{\coderName}{[FIRST CODER'S NAME]}
\newcommand{\coderNameB}{[SECOND CODER'S NAME]}

\makeatletter
\newif\ifanonymous
\if@anonymous \anonymoustrue \else \anonymousfalse \fi
\makeatother

\title{Benchmark Contamination: A Taxonomy Organized by Defeated Mitigation}

\newcommand{\orcid}[1]{\,\href{https://orcid.org/#1}{\textsuperscript{\textsc{iD}}}}

\author{%
 Johanna Angulo\orcid{0009-0005-6965-0604}\thanks{Corresponding author:
 \href{mailto:johanna.angulo@gmail.com}{johanna.angulo@gmail.com}} \\
 Universidad Europea de Madrid \\
 Madrid, Spain \\
 \And
 V\'ictor Yeste\orcid{0000-0002-3660-8347} \\
 Universidad Europea de Valencia \\
 Valencia, Spain \\
 \And
 Hector Espinos-Morato\orcid{0000-0002-4089-1368} \\
 Universidad Europea de Valencia \\
 Valencia, Spain \\
}

\newcommand{\specref}{\citep{angulo2026disclosure}}
\newcommand{\genorefs}{\citep{genoagent2026standard,genoagent2026hard}}

\makeatletter
\renewcommand{\@noticestring}{%
  Preprint. Accepted as a poster at the \@neuripsordinal\ Conference on Neural
  Information Processing Systems (NeurIPS \@neuripsyear). Workshop: TAE
  (Trust-AI-Eval): Can We Trust AI Evaluation? Non-archival.%
}
\makeatother

\begin{document}
\maketitle

\begin{abstract}
\noindent
A benchmark score is a joint property of the model, the harness, the elicitation budget, the sampled population, and contamination status. Leaderboards publish the model and the score, so capability and leakage remain observationally equivalent. Existing taxonomies classify contamination for automated detection, not for the question a reporter faces: given the mitigations already applied, which validity threats remain open? We introduce a taxonomy organized by the mitigation each type defeats (direct, derivative, temporal, distributional, and acquired), spanning training-time and evaluation-time leakage. Holding out a private test set closes the first of these alone. The fifth is acquired \emph{during} evaluation; being a property of one run, it must be recorded with the reported score, not the benchmark release. We operationalize it as a four-field disclosure protocol in which \code{unknown} is a valid entry, released under CC~BY~4.0 with a JSON Schema, a validator, and worked examples. Two coders external to the design team applied a pre-registered instrument to 41 documents. Per-variable linear-weighted $\kappa$ runs from \rKwMin{} to \rKwMax{} (median \rKwMed{}) over \rDocs{} main-pass documents, against a single-coder test--retest ceiling of \rIntraKw{}; pooling raises $\kappa$ to \rKw{} through chance correction, not better agreement, so we report per variable. Of the five variables where no code dominates, three clear the prevalence-robust threshold registered in advance; the two below it are strata reporting and the acquired type. Disagreement concentrates on \emph{when} a variable applies, not on what a document states. Elicitation and regeneration each stay under $25\%$ in every stratum, as registered: elicitation budgets are reported in \rRateFtwo\% of documents, and no document addresses all five types. The contribution is the taxonomy, the score-side artifact that follows, and a pre-registered measurement of instrument reliability and disclosure.
\end{abstract}

\section{Introduction}
\label{sec:intro}
Two students sit the same exam and both score 92\,\%. One reasoned through every
question; the other had prior access to the answer key. From the score alone their performance is indistinguishable: what separates them is the design of the evaluation
and what the examiner discloses. Benchmark evaluation in machine learning faces the same
problem with one added complication: we cannot inspect what the system under test has
seen during training. A reported score is a joint property of at least five factors: the
system, the evaluation harness, the elicitation budget, the population sampled, and contamination status relative to the training corpus. Current practice publishes the model and
the score. Models therefore reach deployment on claimed performance, latency, and cost, not on a documented validity argument, and those claims already inform decisions about patients, defendants, and applicants.

\paragraph{Limitations of existing frameworks.}
Data contamination is well studied, but existing taxonomies are built for detection:
they sort by exposure severity \citep{xu2024survey}, by detection assumptions
\citep{fu2024detection}, or by the structural form of the overlap
\citep{nourbakhsh2026contaminated}. None is organized around the question a reporter faces at
publication: which validity threats to this score have been addressed, and which remain open?
Grouping types by \emph{the mitigation each bypasses} makes one consequence visible:
holding out a private test set closes a single route to inflation. A second gap is that the
contamination literature in NLP/ML venues and the elicitation literature in AI safety venues
rarely cite one another, so a score free of contamination can still be uninterpretable when the
harness and the elicitation budget go unreported. A third is that agentic evaluation introduces
inference-time contamination \citep{han2025searchtime,caisi2025cheating}, which falls outside
surveys that define contamination as training--evaluation overlap.

\paragraph{Contributions.}
\begin{enumerate}\itemsep2pt
\item \textbf{A five-type taxonomy structured by circumvented mitigations}
(\S\ref{sec:taxonomy})---direct, derivative, temporal, distributional, and
acquired. Its central claim is that a private, held-out test set addresses Type~1 alone. Each boundary between types is stated as a decision rule.
\item \textbf{Formalization of inference-time contamination and its reporting implications}
(\S\ref{sec:acquired}). Type~5 is a property of one execution \emph{run}, not of a
benchmark, so benchmark authors cannot certify immunity on an evaluator's behalf. The reporting
artifact must therefore be score-side, not release-side (\S\ref{sec:runproperty}). We add
a tier for the case where the isolation boundary itself is breached.
\item \textbf{A standardized four-field disclosure protocol} (\S\ref{sec:disclosure}). 
The protocol records evaluation strata, elicitation budgets, contamination controls, and
regeneration status. It accepts \code{unknown} as a valid entry and ships with a JSON Schema,
a validator, and worked examples.
\item \textbf{A pre-registered audit of applicability and disclosure}
(\S\ref{sec:instrumentbody}, Appendix~\ref{app:instrument}). A frozen codebook is applied to a
frame of 41 documents (21 released by 7 organizations across two census strata, plus 20
evaluation papers), under an organization-clustered analysis registered before any coding. Two
coders external to the design team applied it. We report inter-coder agreement per variable
alongside disclosure rates, so that reliability is measured. All materials are released.
\end{enumerate}

Four of the five types carry a prior label; Table~\ref{tab:taxonomy} attributes
them. The contribution is the taxonomy: its span across both literatures and the mitigation principle that organizes it. The empirical basis is the audit in \S\ref{sec:instrumentbody}, which measures independent applicability of the types and the completeness of current disclosure against them.

\section{Related Work}
\label{sec:related}
We summarize the core literature underpinning our arguments. Appendix~\ref{app:mapping} details the prior art, artifacts, and quantitative mappings for Table~\ref{tab:taxonomy}.

\textbf{Contamination and its taxonomies.} Because contamination exploitation is setting-dependent \citep{magar2022contamination}, it requires per-benchmark measurement \citep{sainz2023nlp}. Existing taxonomies categorize exposure severity \citep{xu2024survey}, detection assumptions \citep{fu2024detection}, static-to-dynamic shifts \citep{chen2025static}, and item transformations \citep{palavalli2024taxonomy}. While optimized for building detectors, these remain confined to training time. Structuring a taxonomy by the required \emph{remedy} is established in data quality \citep{kim2003dirtydata}; we carry that principle into contamination, and across the training/evaluation boundary that prior taxonomies stop at. Contamination disrupts model ordering, the primary signal for model selection, as inflation is scale-dependent \citep{kocyigit2025overestimation} and triggered even by structurally \emph{similar} training data \citep{tu2024dice,softcontamination2026}, which directly motivates our Type~4.

\paragraph{Evaluation-time contamination.} Existing surveys typically require training--evaluation overlap, omitting inference-time leakage. We integrate \emph{search-time} \citep{han2025searchtime} and \emph{solution contamination} \citep{caisi2025cheating} into a unified framework, adopting the latter's distinction from \emph{grader gaming} (\S\ref{sec:notgrader}).

\paragraph{Reporting and construct validity.} Our protocol extends documentation frameworks like model cards \citep{mitchell2019modelcards}, datasheets \citep{gebru2021datasheets}, and data statements \citep{bender2018datastatements}, but attaches strictly to the \emph{score}---analogous to STREAM \citep{mccaslin2025stream} and Evaluation Cards \citep{ghosh2026evalcards}. The target is construct validity \citep{raji2021everything,bowman2021fix}. Uncertainty estimates are rarely reported \citep{bean2025measuring}, which is why we treat multivariate reporting as a requirement rather than a preference. While evaluation awareness \citep{meinke2024scheming} falls outside our taxonomy, it similarly threatens measurement assumptions, prompting parallel taxonomy-plus-checklist solutions in psychological science \citep{lin2026aimediated}.

\begin{table}[!t]
\centering\footnotesize
\setlength{\tabcolsep}{3.5pt}
\begin{tabularx}{\textwidth}{@{}p{2.75cm} p{2.15cm} X ccccc@{}}
\toprule
\textbf{Artifact} & \textbf{Documents} & \textbf{Contamination treated as} &
\textbf{El.} & \textbf{St.} & \textbf{Rg.} & \textbf{M/R} & \textbf{T5} \\
\midrule
Model cards, datasheets\newline \citep{mitchell2019modelcards,gebru2021datasheets}
 & a model or dataset & out of scope & -- & $\circ$ & -- & -- & -- \\
Contamination Transparency Card \citep{nourbakhsh2026contaminated}
 & a benchmark at release & four tiers of \emph{overlap form} & -- & -- & $\circ$ & -- & -- \\
STREAM \citep{mccaslin2025stream}
 & a score at report & out of scope & $\bullet$ & -- & -- & $\circ$ & -- \\
Evaluation Cards \citep{ghosh2026evalcards}
 & a score at report & one schema line, \emph{before execution} & $\circ$ & $\bullet$ & -- & $\bullet$ & -- \\
\addlinespace[1pt]
\textbf{This form} (\S\ref{sec:disclosure})
 & a score at report & five types by \emph{defeated mitigation} & $\bullet$\rlap{\textsuperscript{*}} & $\bullet$ & $\bullet$ & $\bullet$ & $\bullet$ \\
\bottomrule
\end{tabularx}
\caption{The nearest artifacts against the four fields: elicitation, strata,
regeneration, machine-readability, Type~5. $\bullet$ covered, $\circ$ partial,
-- absent. \textsuperscript{*}by adopting STREAM instead of restating it. The
last column is not a coverage gap: Type~5 is a property of a run
(\S\ref{sec:runproperty}), which no release-side artifact can close.}
\label{tab:priorart}
\end{table}

\paragraph{Positioning.}
\label{sec:priorart}
Table~\ref{tab:priorart} positions our framework against the three nearest artifacts (detailed quantitatively in Appendix~\ref{app:mapping}). The intersection we occupy is a decomposed, per-run contamination record in which \code{unknown} is a valid entry. No release-side artifact can certify inference-time Type~5 contamination (\S\ref{sec:runproperty}). Concurrent work asks whether graders measure task completion \citep{zhu2025abc}; our instrument asks whether the score reflects capability, not exposure.

\section{A Taxonomy Organized by Defeated Mitigation}
\label{sec:taxonomy}

Two foundational claims structure this taxonomy: a held-out private test set mitigates solely Type~1, and whereas Types 1--4 represent passive data leakage \emph{into} the model, Type~5 constitutes active data acquisition \emph{by} the model. 
Both are relative to the construct: prior access is not contamination in itself but becomes so when it lets the system reach the score by a route other than the capability
claimed.
Figure~\ref{fig:taxonomy} shows the paths and Table~\ref{tab:taxonomy} summarizes
them.

\textbf{Why five.} A score is inflated by contamination when information
correlated with the answer reaches the system by a route other than the
capability claimed. Two questions partition the routes: does the information
arrive through training or during the run, and, if through training, what is it
about? It can be about the item itself, in any form (Type~1); about the evidence
the item was written from (Type~2); about the event that resolved the item after
it was written (Type~3); or about no single item but the population it was drawn
from (Type~4). Arrival during the run is Type~5, whatever the channel. Item,
antecedent, consequent and population exhaust what training-time information can
be about relative to one item. The failures we know of that fall outside this
space (grader gaming and evaluation awareness) move no answer information and
are excluded by definition (\S\ref{sec:notgrader}). The argument covers the
space it defines, not every future failure. The unit identifies a type; the
defeated mitigation orders the types. Each type survives at least the
characteristic remedy of the one before it (Table~\ref{tab:taxonomy}), so the
chain is cumulative rather than strictly nested.

\begin{figure}[!t]
\centering
\begin{tikzpicture}[x=1cm,y=1cm]

\node[cdbox, text width=2.55cm] (s1) at (1.55,3.25)
 {\textbf{Benchmark items}\\[1pt] items {+} labels, crawled};
\node[cdbox, text width=2.55cm] (s2) at (4.75,3.25)
 {\textbf{Source documents}\\[1pt] what the items were built from};
\node[cdbox, text width=2.55cm] (s3) at (7.95,3.25)
 {\textbf{Resolved outcomes}\\[1pt] answers published before the cutoff};
\node[cdbox, text width=2.55cm] (s4) at (11.15,3.25)
 {\textbf{Item patterns}\\[1pt] the template, not the instance};

\node[cdmodel, minimum width=9.6cm, minimum height=0.72cm] (m) at (6.35,1.62)
 {System under test};

\foreach \s/\x/\n in {s1/3.0/1, s2/5.1/2, s3/7.6/3, s4/9.7/4}{
 \draw[cdarr] (\s.south) -- (\x,1.98);
 \node[cdtag] at ($(\s.south)!0.55!(\x,1.91)$) {\n};
}

\draw[dashed, draw=cdacc, line width=0.5pt] (0.15,0.62) -- (12.55,0.62);
\node[font=\scriptsize\itshape, text=cdacc, anchor=east] at (12.55,0.90)
 {isolation boundary};

\node[cdacq, text width=2.7cm] (a) at (2.35,-0.35)
 {\textbf{5a} environment\\[1pt] repo history, config files};
\node[cdacq, text width=2.7cm] (b) at (6.35,-0.35)
 {\textbf{5b} open web\\[1pt] the benchmark, hosted with labels};
\node[cdacq, text width=2.7cm] (c) at (10.35,-0.35)
 {\textbf{5c} adversarial\\[1pt] the boundary itself is defeated};
\foreach \t/\x in {a/2.35, b/6.35, c/10.35}{
 \draw[cdarr5] (\x,1.19) -- (\t.north);
}
\node[cdtag5] at (6.35,0.62) {5};

\node[cdband, anchor=west] at (0.15,4.25) {training time \ \textbf{---} \ the data reaches the model};
\node[cdband, anchor=west] at (0.15,-1.32) {evaluation time \ \textbf{---} \ the model reaches the data};
\end{tikzpicture}
\caption{The five types and the paths they travel; the arrow direction is the
distinction that matters. At training time four paths run \emph{into} the system.
At evaluation time one runs \emph{outward}, across the isolation boundary, to an
answer key already in the environment (5a), reachable on the web (5b), or
reachable only once the boundary is defeated (5c).}
\label{fig:taxonomy}
\end{figure}

\begin{table}[!t]
\centering\footnotesize
\setlength{\tabcolsep}{3pt}
\begin{tabularx}{\textwidth}{@{}l >{\raggedright\arraybackslash}p{1.55cm}
 >{\raggedright\arraybackslash}X >{\raggedright\arraybackslash}X
 >{\raggedright\arraybackslash}X@{}}
\toprule
\textbf{Type} & \textbf{Unit} & \textbf{Closed by} & \textbf{Survives (prior type's remedy)} & \textbf{Documented as, and where} \\
\midrule
1 Direct & The item, in any form & holding the item out; item and label decontamination & --- & exact, verbatim, label-level or rephrased-item contamination \citep{magar2022contamination,xu2024survey,deng2024investigating,palavalli2024taxonomy,yang2023rephrased,balloccu2024leak} \\
2 Derivative & The source document & provenance control; items needing integration across sources & holding out (Type~1's remedy) & no prior label; the nearest term, indirect contamination \citep{fu2024detection}, runs the other way and falls in Type~1 (Appendix~\ref{app:ctc}) \\
3 Temporal & The resolved outcome & items on phenomena unresolved at the cutoff; temporal splits & provenance control: the sources predate the resolution & task contamination; temporal analysis \citep{li2024task,testoftime2025} \\
4 Distributional & The pattern & construct-preserving perturbation; regeneration from a published generator & post-cutoff construction: new item, old template & in-distribution or soft contamination \citep{tu2024dice,dekoninck2024constat,softcontamination2026} \\
5 Acquired & The evaluation run & sanitization, isolation, transcript review; at 5c, boundary monitoring (detective only) & every training-side remedy, including regeneration; at 5c, network isolation itself & search-time; solution contamination \citep{han2025searchtime,caisi2025cheating,wang2026stcdeepresearch,marchand2026sandboxescape} \\
\bottomrule
\end{tabularx}
\caption{The five types, organized by unit and by the mitigation each defeats:
each row survives at least the characteristic remedy of the row above, and only
Type~1 is closed by holding the benchmark out. The last column is the paper's own
novelty audit: every type but Type~2 has a prior label, across three literatures
that largely do not cite one another, and row~5's sources are disjoint from rows
1--4's.}
\label{tab:taxonomy}
\end{table}

\subsection{Types 1--4: The Data Reaches the Model}

\textbf{Type 1 --- Direct.} \emph{Benchmark items and their labels exist in the model's pretraining or post-training corpus.} This occurs when public benchmarks are scraped into training crawls. This is the classic scenario $n$-gram overlap metrics target, which strictly held-out private sets prevent by construction---provided the dataset remains isolated.
\textbf{Why detection still fails:} black-box probing \citep{deng2024investigating} cannot verify the absence of overlap in closed-weight models, so for those models the truthful entry is often \code{unknown}.
\textbf{Where Type~1 ends.} The line with Type~2 is drawn by one question: would
holding the benchmark out have prevented the exposure? If the item reached the
corpus in any form (verbatim, partially paraphrased, translated, or
reformatted), it would have, and the case is Type~1 however lexical checks fare
against it: rephrased items evade $n$-gram filters, yet training a 13B model on
them lifts performance to GPT-4 levels \citep{yang2023rephrased}. If what
reached the corpus is the material the item was written from, the case is
Type~2. Partial paraphrase of the item is therefore Type~1;
Appendix~\ref{app:ctc} maps the tiers of the Contamination Transparency Card
onto the five types. The audit's \code{t2} variable codes attention to source
material only (Table~\ref{tab:instrument}), so no reported result depends on
where paraphrase falls.

\textbf{Type 2 --- Derivative.} \emph{The benchmark never leaked, but its source material is in the training corpus.} For example, a private clinical-genomics benchmark built from published case reports is compromised because those underlying reports exist in standard web crawls. The model succeeds by retrieving source-documented associations instead of recognizing the test item. \textbf{Why held-out sets miss it:} Holding out a benchmark built on public evidence is ineffective; it is contaminated upon creation. This is endemic to expert domains (e.g., clinical genomics, law, cybersecurity) where ground truth must be publicly documented before evaluation use.

\textbf{Type 3 --- Temporal.} \emph{The model's training cutoff postdates the tested phenomenon, reducing reasoning or forecasting to mere recall.} If an item requires predicting a causal gene based on historical evidence, but the model's training cutoff postdates the gene's actual discovery, the model simply retrieves the known fact. This structural flaw frequently compromises forecasting tasks and historical code benchmarks.
\textbf{Why overlap checks miss it:} Because test items may be newly authored, no lexical overlap exists. Leakage occurs at the factual rather than lexical level: the answer is already internalized by the model as parametric world knowledge.

\textbf{Type 4 --- Distributional.} \emph{The items are new and held out, but their structural pattern is so well represented in training that the model bypasses the intended reasoning.} For example, models scoring highly on newly authored word problems often fail when entities are renamed, numbers altered, or logically inert clauses introduced \citep{mirzadeh2024gsmsymbolic}. No direct data leaked; the item is new, but the template is not. \textbf{Why overlap methods miss it:} the unit of leakage is the distribution, not the item. Types 1--3 inflate the score; Type~4 changes what the score \emph{measures} and leaves no inflation signal. At this point contamination stops being data hygiene and becomes a construct-validity failure.

\textbf{When to call it Type~4.} We set no universal threshold, because the line
is relative to the construct; we set a procedure. The reporter fixes, before
scoring, a family of transformations that preserve the construct (renamed
entities, altered numbers, inert clauses, regenerated instances) and reports the
score on canonical and transformed items with intervals. Type~4 is
\code{controlled} only when that delta is reported. A delta inside chance
variation is no evidence of Type~4; one outside it is evidence, not proof, and
the reader judges it on evidence, not on assurance.

\subsection{Type 5 --- Acquired}
\label{sec:acquired}

\textbf{Definition.} The model acquires the ground-truth answers \emph{during} the evaluation through retrieval, tool use, filesystem access, or, in extreme cases, by breaching the isolation boundaries of the evaluation environment. Unlike Types 1--4, this contamination is a property of the system's dynamic behavior rather than a static dataset--corpus relationship. This distinction warrants a separate category and necessitates the reporting protocols outlined in \S\ref{sec:runproperty}.

\textbf{Three levels, by the boundary crossed.}
\begin{itemize}\itemsep3pt
 \item \textbf{5a --- Environmental.} The answer resides within the evaluation environment
 itself: version-control histories holding fixing commits, solution strings in task
 configuration files, residual artifacts in containers
 \citep{caisi2025cheating,swebench2025issue465}. No external network access is required, and
 restricting access to future commits demonstrably reduces measured performance
 \citep{trajectories2026}.

 \item \textbf{5b --- Retrieval.} The agent searches the open web during evaluation and
 retrieves the benchmark with its labels \citep{han2025searchtime,wang2026stcdeepresearch}.
 This bypasses decontamination and training-cutoff controls alike, since acquisition occurs at
 inference time: a benchmark released \emph{after} a model's cutoff remains vulnerable, and a
 held-out set is compromised the moment it becomes reachable online. Network isolation is the
 main control available.

 \item \textbf{5c --- Adversarial.} The agent defeats the evaluation's isolation boundaries.
 In a July 2026 incident, models evaluated on a cyber-capability benchmark
 \citep{wang2026exploitgym} escaped their sandbox and chained exploits into third-party
 infrastructure
 \citep{openai2026hfincident,huggingface2026disclosure,metr2026hfincident}; the sequence
 is in Appendix~\ref{app:5c}. Level 5c defeats \emph{network isolation}, the control
 recommended for 5a and 5b. Preventive controls remain open; detective measures are in
 Appendix~\ref{app:detection}. The underlying capability is measurable
 \citep{marchand2026sandboxescape}.
\end{itemize}

\textbf{Scope of the anchor case, and relation to prior work.} The July 2026 incident is one disclosed event, run with safety configurations deliberately relaxed, and supports no inference about frequency or intent. The fullest accounts sharpen what it anchors. Whether the breach of the host reached an answer key is disputed---the operator's first disclosure says it did \citep{openai2026hfincident}, the independent investigation finds the attack aimed at the scorer \citep{metr2026hfincident}---so what it demonstrates is that isolation can fail against a capable system. Both agree the same runs acquired by another route, deriving each task's answer during the run \citep{metr2026hfincident,openai2026roadahead}. Acquisition, grader gaming and a boundary breach arrived entangled in one incident, which is the argument for recording them separately (Appendix~\ref{app:5c}). The phenomenon is known in agentic evaluation as search-time \citep{han2025searchtime} and solution contamination \citep{caisi2025cheating}. Our formulation categorizes the invariant, not the channel, since channels multiply as tool access expands; it adds a tier for breached isolation and places these behaviors in one taxonomy with training-time contamination, which prior frameworks do not. Table~\ref{tab:taxonomy} makes the comparison type by type.

\subsubsection{Why Type 5 changes what a disclosure can certify}
\label{sec:runproperty}

Types 1--4 are properties of a dataset--corpus relationship: assessable once at release and stable across later uses. Type~5 is a property of one model, in one harness, at one time. A benchmark uncontaminated with respect to Types 1--4 can still be compromised depending on whether a \emph{given run} permitted network access and whether \emph{that model} retrieved the solutions. Three consequences follow, each reflected in the reporting format.

\begin{enumerate}\itemsep2pt
 \item \textbf{Type~5 must be certified per run, not per benchmark.} Benchmark authors cannot preemptively guarantee immunity to Type~5 contamination on behalf of practitioners. This necessitates a score-side, rather than release-side, reporting artifact (\S\ref{sec:priorart}).
 \item \textbf{Benchmark validity degrades as agentic capabilities increase}, irrespective of the benchmark's original design. While documented instances predominantly involve highly agentic models, this correlation suffers from selection bias: only agentic evaluations generate the explicit execution traces necessary to observe data acquisition. The vulnerability comes from the mechanism itself: a channel that only an autonomous system can navigate.
 \item \textbf{Transcript review is required.} At Levels 5a and 5b the evidence sits in the execution trace: version-control queries, retrieved URLs, reasoning steps showing the answer was found, not derived. At Level 5c the trace is not enough, because a system that breached the boundary is no witness that it held: in the one investigated case the transcripts were themselves altered \citep{metr2026hfincident}. Boundary monitoring is therefore recorded separately (\S\ref{sec:disclosure}).
\end{enumerate}

Dispositional controls, such as system prompts telling the model not to seek answers, are excluded: a certification of evaluation integrity cannot rest on the compliance of the system under test.

\paragraph{The boundary with grader gaming, and why it is a boundary.}
\label{sec:notgrader}
One question separates them: does the intervention change what the system knows about the \emph{answer}, or what the scorer does with an answer already fixed? The first is Type~5; the second is grader gaming, addressed by reward and scorer design \citep{caisi2025cheating,zhu2025abc} and outside this taxonomy. They are kept apart because they defeat different mitigations (isolation and transcript review against one, scorer design against the other), so merging them would say that something is wrong without saying which remedy is owed. They nonetheless arrive together, as they did in the July 2026 runs \citep{chen2026publicscore,metr2026hfincident}. A disclosure should therefore record what a breach was \emph{for} in the Type~5 note; \code{controlled} on Type~5 is no claim about the grader, and scorer integrity needs a record of its own that this form does not supply. Type~5 assumes nothing about intent: saying a model ``cheated'' names an outcome, not a motive.

\section{The Disclosure Form}
\label{sec:disclosure}

The taxonomy necessitates a four-field reporting artifact per published score (see Appendix~\ref{app:fields} for fallback disclosures). 

\textbf{Rationale.} \emph{Contamination controls} captures Types 1--5. \emph{Regeneration} remains distinct because publishing generation procedures mitigates Types 1, 3, and 4 simultaneously, characterizing the instrument rather than the run. \emph{Strata} and \emph{elicitation budget} are prerequisites: uncontaminated scores are uninterpretable without specified populations or compute budgets. To minimize friction, statistical variance and run counts are delegated to interoperable frameworks like STREAM \citep{mccaslin2025stream}. The form is a contamination module meant to travel with a fuller score record: F3 and F4 carry what existing artifacts do not record, while F1 and F2 are the minimum without which F3 is uninterpretable, and either can be satisfied by reference to an Evaluation Card \citep{ghosh2026evalcards} or STREAM record that already states it.

\textbf{Strata reported.} An aggregate hides systematic failure on long-tail strata such as rare diseases or low-resource languages. Disclosing minority-subset performance imposes minimal overhead, as per-stratum metrics are routinely computed during evaluation pipelines. A complete entry names the stratification variable and its source (published with the benchmark, fixed by the reporter before scoring, or defined post hoc) and gives each stratum's $n$ and score with an interval; an aggregate-only report says so and why (Appendix~\ref{app:fields}).

\textbf{Elicitation budget.} Capability scores reflect both the model and its harness. Because tuning deltas often exceed generational gaps, claims like ``the model cannot execute X'' may merely indicate insufficient compute. Thus, inability claims without explicit budgets characterize harness limits, not model limits (Appendix~\ref{app:elicitation}). A complete entry carries the five elements the audit records separately (Appendix~\ref{app:instrument}): an identified harness---named, pinned to a public version, or fully specified as a scaffold---its version, the budget, the number of attempts, and how attempts were resolved.

\textbf{Contamination controls.} Reporters assign \code{controlled}, \code{not\_controlled}, \code{unknown}, or \code{n/a} per type. Decontaminating a test set satisfies only Type~1. Type~5 requires tracking network access, sanitization, transcript review, and active boundary monitoring. Boundary monitoring is separated because a compromised system's trace cannot reliably testify to its own containment (Level 5c). Type~5 is reported as one status, so that disclosures compare at a glance; its three levels determine it. A level is \code{n/a} when its channel is absent: 5a with no persistent environment, 5b with no network path, and all three when the system has no channel outside its weights. Type~5 is \code{controlled} only if every applicable level is, \code{not\_controlled} if any applicable level is not, and \code{unknown} otherwise. Because nothing prevents 5c, it counts as controlled only when boundary monitoring is recorded, and counts as detection, not prevention. This applicability rule is the decision procedure the audit found missing (\S\ref{sec:instrumentbody}).

\textbf{Regeneration.} Static artifacts leak into training corpora over time. Publishing the \emph{generation procedure} shifts the epistemic burden from trusting opaque claims to enabling independent regeneration. Continuously refreshed benchmarks \citep{white2025livebench} practically approximate this principle.

\textbf{Why \code{unknown} is needed.} No independent practitioner can verify a closed-weight pretraining corpus, and demanding certainty penalizes the reporter who is candid. Permitting \code{unknown} separates a question that cannot be answered from one that was never asked, and it is what makes the form adoptable. An \code{unknown} entry should state its reason in the note: \emph{unverifiable} (no reporter could know, as with a self-reported cutoff), \emph{not assessed} (checkable but not checked), or \emph{withheld} (known but confidential). The reasons separate structural limits from neglect, and they fall unevenly: Types~4 and~5 are establishable by the reporter alone (perturbation needs no corpus, and transcripts are the reporter's own), so \code{unknown} there is usually \emph{not assessed}, while Types~1--3 often need corpus access or a verified cutoff only the provider holds. Where \code{unknown} accumulates therefore marks where provider attestation is needed, which is itself information a reader can use.

The formal specification \specref{} is released under CC~BY~4.0 with field definitions, templates, a JSON Schema, worked examples \genorefs{} (Figure~\ref{fig:form}), and a validator that flags weak disclosures. Machine-readability lets continuous-integration pipelines catch incomplete reporting without manual review (Appendix~\ref{app:artifacts}).


\begin{figure}[!t]
\centering
\begin{minipage}[t]{0.78\textwidth}
\centering
\begin{tikzpicture}
\node[draw=cdline, rounded corners=2.5pt, inner sep=5pt, fill=white, align=left]{%
\scriptsize
\begin{tabular}{@{}>{\raggedright\arraybackslash}p{0.72cm}@{\ }>{\raggedright\arraybackslash}p{6.05cm}@{}}
\multicolumn{2}{@{}l@{}}{\textbf{\footnotesize Disclosure record} \;\; \textit{agentic clinical-genomics}}\\[1pt]
\multicolumn{2}{@{}l@{}}{\rule{0pt}{5pt}\hrulefill}\\[2pt]
\textbf{F1} & \textbf{strata:} four published strata of the cohort, each with
 its $n$ and interval \\[2pt]
\textbf{F2} & \textbf{elicitation:} harness pinned to a commit; one attempt;
 temperature~0; retrieval disabled during scoring \\[3pt]
\textbf{F3} & \textbf{contamination controls} \\[1pt]
\textit{t1} & \code{controlled} --- direct: scored split not published with
 labels; canary string embedded \\
\textit{t2} & \code{not\_controlled} --- derivative: items derive from indexed case reports \\
\textit{t3} & \code{unknown} --- temporal: model cutoffs are self-reported and unverifiable \\
\textit{t4} & \code{not\_controlled} --- distributional: by design; the delta
 against the hard variant is the measurement \\
\textit{t5} & \code{controlled} --- acquired: network off, transcripts
 reviewed; \emph{boundary monitoring: none} \\[3pt]
\textbf{F4} & \textbf{regeneration:} artifact only; no generator published \\[3pt]
\end{tabular}};
\end{tikzpicture}\\[2pt]
\end{minipage}
\caption{An abridged disclosure record adapted from worked examples in the formal specification \specref{} (scores omitted). Declaring \code{unknown} and \code{not\_controlled} shows what the instrument is for: recording a limitation costs the reporter little and tells the reader more than an omitted field. The validator flags this record, because a \code{controlled} claim for Type~5 without boundary monitoring is an assumption rather than a control. Under the level rule of \S\ref{sec:disclosure}, this record's Type~5 entry resolves to \code{unknown}, since level~5c has no boundary evidence.}
\label{fig:form}
\end{figure}

\section{Can the Taxonomy Be Applied Independently?}
\label{sec:instrumentbody}
\textbf{What the coders did.} Each coder received one document at a time---a
system card, a benchmark paper, or a third-party evaluation report---and a manual
generated from the frozen codebook. They first fixed
one \emph{focal evaluation} mechanically: the first capability score in the body
text, kept even when a better-documented evaluation appears later, so that the
study measures typical rather than best practice. For each of the eight variables
they then searched the document with listed terms and recorded \code{2}
(stated), \code{1} (mentioned without the required detail), \code{0} (searched
and absent) or \code{NA} (cannot apply), with a page locator for every non-zero
code. In the manual's worked example, a system card reports GPQA Diamond with
per-subject accuracy, ``single attempt'', an unnamed ``standard internal
evaluation stack'', a training cutoff never related to the items, and a global
``network access disabled''. It codes \code{2} on strata, \code{1} on elicitation
(no harness a reader could look up), \code{1} on temporal, \code{2} on acquired
and \code{0} elsewhere. Two coders external to the design team worked
independently, in different seeded orders. \code{NA} counts as a code that can be
disagreed on, not a dropped cell, and because the coders calibrated on the pilot,
primary $\kappa$ uses the independent main pass alone. Agreement was computed and
saved before an author who coded nothing adjudicated the disputed cells used for
the disclosure rates. The design, statistics and decision rules were registered
before coding (Appendix~\ref{app:instrument}).

\begin{table}[!ht]
\centering\footnotesize
\setlength{\tabcolsep}{5pt}
\begin{tabular}{@{}l r r r l r r r r@{}}
\toprule
& \multicolumn{5}{c}{\textbf{Agreement} (two coders, main pass)}
& \multicolumn{3}{c}{\textbf{Disclosure} (adjudicated)} \\
\cmidrule (lr){2-6}\cmidrule (l){7-9}
\textbf{Variable} & $n$ & Raw & $\kappa_w$ & 95\% CI & AC2 &
\code{2} & \code{1}+\code{2} & $k$ \\
\midrule
F1 Strata & 29 & 45\% & 0.18 & [-0.12,~0.47] & 0.43 & 39\% & 58\% & 26 \\
F2 Elicitation & 29 & 48\% & 0.10 & [-0.10,~0.32] & 0.63 & 13\% & 92\% & 26 \\
F3 t1 Direct & 29 & 69\% & 0.33 & [\phantom{-}0.02,~0.63] & 0.68 & 5\% & 21\% & 26 \\
F3 t2 Derivative & 29 & 79\% & 0.00 & [\phantom{-}0.00,~0.00] & 0.89 & 0\% & 11\% & 26 \\
F3 t3 Temporal & 29 & 83\% & 0.31 & [-0.06,~0.76] & 0.88 & 8\% & 16\% & 26 \\
F3 t4 Distributional & 29 & 79\% & 0.00 & [\phantom{-}0.00,~0.00] & 0.77 & 8\% & 11\% & 26 \\
F3 t5 Acquired & 29 & 55\% & 0.24 & [-0.01,~0.48] & 0.50 & 5\% & 38\% & 11 \\
F4 Regeneration & 29 & 59\% & 0.35 & [\phantom{-}0.11,~0.58] & 0.69 & 11\% & 50\% & 26 \\
\midrule
Pooled$^{\dagger}$ & \rDocs & \rRaw\% & \rKw & [\rKwLo,~\rKwHi] & 0.66 & & & \rOrgs \\
\bottomrule
\end{tabular}
\caption{Agreement and disclosure per coded variable (\code{score.py} output). $\kappa_w$: linear-weighted Kappa with a 95\% bootstrap interval; AC2: Gwet's prevalence-robust companion under the same weights, reported because $\kappa$ collapses under class skew. F1 and t5 fall below the registered $0.6$ AC2 threshold. \code{2}: full reporting; \code{1}+\code{2}: any reporting; $k$: contributing clusters. Agreement excludes the pilot; disclosure uses all $38$ included documents. $^{\dagger}$Not pre-registered; why the pooled value exceeds every per-variable value is explained in Appendix~\ref{app:instrument}.}
\label{tab:results}
\end{table}

\textbf{Agreement.} On the \rDocs\ main-pass documents, pooled linear-weighted
$\kappa$ is \rKw\ (bootstrap 95\%~CI [\rKwLo,~\rKwHi]), with \rRaw\% raw
agreement and \rACone\ Gwet's AC1. Per-variable $\kappa$ runs from \rKwMin\
(\rKwMinVar) to \rKwMax\ (\rKwMaxVar) (Table~\ref{tab:results}). These figures
are read against the measured single-coder test--retest ceiling
$\kappa_w=\rIntraKw$, not against verbal bands. Of the $35{\times}8$ cells both
coders coded, \rAdjud\ went to adjudication; the coders named the same focal
benchmark on $25$ of the $35$ documents (Appendix~\ref{app:instrument}).

\textbf{Prevalence versus unreliability.} The registration anticipated $\kappa$
collapse under skew and fixed a divergence rule against a weighted
prevalence-robust companion. That rule accounts for t2 ($0.00$ vs.\ AC2 $0.89$),
t4 ($0.00$ vs.\ $0.77$), and t3 ($0.31$ vs.\ $0.88$): the code is almost always
\code{0}. Those categories are rare, not unusable. Two variables are not
explained by rarity. Strata reporting has AC2 $=0.43$ and acquired contamination
(Type 5) AC2 $=0.50$, both below the registered $0.6$ threshold for a non-rare category;
here the prevalence-robust measure and $\kappa$ agree that agreement is poor.

\textbf{Applicability, not reading.} Both large splits are threshold questions:
what counts as an identified harness for elicitation budget (\code{f2}), and when
acquired contamination (\code{t5}) applies. A clarification issued to both coders
in the same words moved both without converging them: ten of the eighteen
remaining \code{t5} disagreements turn on whether the variable applies at all,
in both directions, and adjudication found a channel in each
(Appendix~\ref{app:instrument}).

\textbf{Disclosure.} \rRateFone\% of documents report per-stratum scores,
\rRateFtwo\% elicitation budgets, and \rRateFfour\% instrument regeneration
status. Only \rRateTone\% address direct contamination and \rRateTfive\%
acquired contamination---one third-party pilot document, without which the rate
is $0\%$---and none addresses all five types
(Table~\ref{tab:results}). The \code{t5} rate uses the $21$ of $38$ documents
with a channel outside the weights. The \code{1}+\code{2} column shows how far mentions outrun complete
reports. Every rate carries its cluster count: $38$ documents over \rOrgs\
clusters, seven organizations covering $19$. The registered tie-break was never
reached, and the sensitivity bands are in Appendix~\ref{app:instrument}.

\textbf{Readability and disclosure are distinct.} System cards are the easiest
stratum to code and disclose the least; third-party reports are the hardest to
code and are not the most forthcoming, which benchmark papers are on seven of
the eight variables. Coding difficulty therefore does not track disclosure. A
system card covers dozens of evaluations under a page budget, so per-evaluation
method detail is structurally absent rather than withheld
(Table~\ref{tab:strata}). Of the three registered hypotheses only the first is
supported---elicitation and regeneration each stay under $25\%$ in every
stratum---and the primary contrast runs opposite the predicted direction on an
interval excluding zero (Appendix~\ref{app:instrument}).

\textbf{What the agreement licenses.} Three readings follow. \emph{Reliable now:}
among the variables where no code dominates, elicitation, direct contamination
and regeneration clear the registered AC2 threshold ($0.63$, $0.68$, $0.69$) and
can be coded by outsiders from the manual. \emph{Untested rather than reliable:}
agreement on derivative, temporal and distributional contamination is agreement
on absence, because presence is rare, so the audit cannot yet show that coders
agree when these controls are reported. \emph{Not yet reliable:} strata reporting
and acquired contamination, both traced to threshold rules stated in prose. For
the form itself, coding a document is harder than completing a record, because
the author states the harness and budget instead of inferring them. The Type~5
applicability judgment survives the change of task, however, since an author
must also decide whether a channel exists; that is why the level rule of
\S\ref{sec:disclosure} belongs to the form and not only to the manual. The audit
applies the taxonomy to $38$ real documents across three genres. It tests whether
disclosures can be coded, not whether the five types cover every contamination
case (\S\ref{sec:limitations}).

\section{Limitations}
\label{sec:limitations}
\textbf{Scope.} The protocol is a reporting standard, not a validity guarantee;
construct validity remains a separate prerequisite. It records claimed
mitigations
instead of detecting contamination. Disclosures are self-reported and cannot be
verified externally: the form makes an absent claim conspicuous, not a false one
impossible. The audit in \S\ref{sec:instrumentbody} tests independent
applicability of the taxonomy, not the truth of the disclosures.

\textbf{Sample and inference.} Agreement rests on one coder pair and \rDocs\
documents. Under the pre-registered criterion, strata reporting and the acquired
type are not yet reliably applicable by coders who did not write the manual. One
pair cannot separate an unclear construct from two divergent readers; what
points to the manual is where the disagreement sits (at stated thresholds, in
both directions) rather than spread across cells. The test--retest is weaker than it looks: the registration
asked both coders to recode five documents and one did, so
$\kappa_w=\rIntraKw$ bounds that coder's own noise instead of establishing the
spread as a property of the manual. The agreement figures are in turn an upper bound in the registered
sense: $\kappa$ is
computed only on the $35$ documents both coders included, and the three they
disputed are the hardest in the frame. With $21$ of $41$
documents from seven organizations sharing house templates, the design is sized
for the agreement question rather than for cross-institutional rate comparison.
The rates are a first application, not a field-wide census
\citep{bean2025measuring}; what they add is an agreement statistic
\citep{reed2025modelreports}.

\textbf{Coding scheme versus author protocol.} The audit evaluates the taxonomy
as a coding scheme: a third party infers, from a document not written to the
scheme, whether a condition was met. The protocol of \S\ref{sec:disclosure} is
completed by the party who ran the evaluation and who states the harness and budget instead of inferring them. The partial-versus-full judgment that
Table~\ref{tab:collapse} identifies as the main locus of disagreement has no
analogue in that reporting task. The audit therefore does not measure protocol
usability, which requires author-completion and reader-comprehension studies
still absent for artifacts in
this literature \citep{mitchell2019modelcards,gebru2021datasheets}. It does
bound the gap the protocol addresses: across the $38$ included documents and eight
variables, $11\%$ of cells record a field as fully reported and $60\%$ as absent.
A protocol that admits \code{unknown} would make explicit what is at present
simply missing.

\textbf{Coverage and next steps.} Beyond the audit's $38$ documents and the
literature behind Table~\ref{tab:taxonomy}, the taxonomy and form have not been
stress-tested against documented contamination cases on a broad benchmark set.
That study, with independent annotators and a registered protocol, is the next
step. It will run alongside an author-completion study of the form and a re-test
of the boundary rules in \S\ref{sec:taxonomy}--\S\ref{sec:disclosure}, which were
written after the audit's disagreements were seen and are untested.

\textbf{Taxonomy boundaries.} Post-training, instruction-tuning, and
distillation leakage are subsumed into Types~1 and~2 despite distinct
behavioral signatures; multimodal contamination is excluded. Two boundaries are
operational. Type~4 and construct validity are functionally similar as
claims---over-represented training patterns versus failure to measure the
intended construct---and are separated by taking the \emph{training
distribution} as the causal factor. The Type~5a/Type~1 boundary blurs when the
evaluation container is itself the artifact. Level~5c rests on a single
disclosed incident and establishes possibility, not frequency
(Appendix~\ref{app:5c}).

\section{Conclusion}
Benchmark contamination is treated as one failure with one remedy. It is at
least five, and a private held-out test set addresses one of them. Acquired
contamination is a property of a run: it cannot be certified at benchmark
release, and the party publishing the score is the party positioned to disclose
it.

Two external coders working from a frozen manual reached per-variable
$\kappa_w$ between \rKwMin\ and \rKwMax\ (median \rKwMed) against a
single-coder test--retest ceiling of \rIntraKw. Disagreement concentrates on
applicability: ten of the eighteen residual Type~5 splits turn on whether the
variable applies at all rather than on what a document states. No document
addresses all five contamination types; elicitation budgets are reported in
\rRateFtwo\% of documents. The taxonomy locates a near-total disclosure gap and
identifies which applicability boundaries need tighter specification. The form
now states those boundaries as rules (where Type~1 ends, when Type~4 counts as
controlled, when each Type~5 level applies), and testing them is the subject of
the next registered study.

The instrument is released so the measurement can be repeated, disputed, or
extended. A published score should arrive with more than a single number;
\code{unknown} is a valid answer.

\newpage
\bibliographystyle{plainnat}
\bibliography{references}

\section*{Ethics Statement}

This paper analyzes published documents and released artifacts: no personal data.
Two people code the audit, both external to the design team and neither an author.
Neither is a study participant --- both apply a released manual to public
documents, nothing is recorded about them beyond the codes they assign, and the
released sheets carry role labels, not names --- so no ethics review is engaged;
their agreement to code, and to be credited by role label only, is on file. The 2026 incident is discussed only from the parties'
public disclosures and a published independent investigation of them
\citep{metr2026hfincident}, without operational detail.

The impact runs both ways. A cheap record should better inform selection and
makes a missing control visible where silence now reads as competence. The risk
is that disclosure substitutes for control: five types offer five places to write
\code{controlled} without evidence. The validator flags valid-but-weak records, but the form is
unverifiable, and a standard hardened into ritual would be worse than none.

\section*{Availability}

The specification (v1.1), taxonomy, Markdown and YAML templates, JSON Schema,
validator, worked examples, audit instrument and a full prior-art analysis are
released under CC~BY~4.0 at
\url{https://github.com/Jangulo7/contamination-disclosure-paper} \specref{} and
archived at \url{https://doi.org/10.5281/zenodo.22182947}. The registration is
public on OSF at \url{https://doi.org/10.17605/OSF.IO/Q6W9A}. An earlier four-type
version of the taxonomy was presented as a talk and archived separately
\citep{angulo2026talk}; that version is designated v1.0 and the five-type version
described here is v1.1, so the archived artifact and the released specification
cannot be confused.

\begin{ack}
\ifcodersnamed
\coderName\ and \coderNameB\ carried out the independent coding pass, each
\else
Two independent coders (\code{R1} and \code{R2}, the role labels on their
deposited sheets) carried out the coding pass, each
\fi
applying the frozen codebook to the full frame, and neither having helped
design the taxonomy or the manual. That someone other than its authors can use
the taxonomy rests on their coding, not ours.

This work received no external funding.
\end{ack}



\appendix

\section{The released instrument}
\label{app:instrument}
\label{app:artifacts}
This appendix condenses the frozen coding manual; Figure~\ref{fig:instrument}
gives the audit end to end. The full manual, sampling frame, pre-registration,
and analysis scripts are in the supplementary materials.

\begin{figure}[!ht]
\centering
\begin{tikzpicture}[x=1cm,y=1cm,
 hstep/.style={cdstep, text width=3.05cm, font=\scriptsize}]
\node[hstep] (f) at (0,0)
 {\textbf{Frame: 41 documents}\\ 21 from 7 organizations; {+}\,20 evaluation papers};
\node[hstep] (r) at (3.55,0)
 {\textbf{Focal-evaluation rule}\\ one document {=} one evaluation, chosen mechanically};
\node[hstep] (c) at (7.1,0)
 {\textbf{8 variables} $\times$ \code{2/1/0/NA}\\ two external coders, seeded orders};
\node[hstep] (a) at (10.65,0)
 {\textbf{$\kappa_w$}, bootstrap CI, AC2; rates as ``$k$ organizations, $n$ documents''};
\foreach \i/\j in {f/r, r/c, c/a}{\draw[cdarr] (\i.east) -- (\j.west);}
\end{tikzpicture}
\caption{The audit end to end: frame, focal-evaluation rule, coding design, and the statistics \code{score.py} computes from the two sheets.}
\label{fig:instrument}
\end{figure}

\paragraph{Unit and scale.}
The unit is a document, coded against one \emph{focal evaluation}: the first
capability benchmark score in the body text, selected mechanically so that a
system card with fifty evaluations remains comparable to a third-party report
with one. Every variable is coded \code{2} (reported), \code{1} (partial),
\code{0} (absent), or \code{NA} (not applicable). \code{NA} means the variable
cannot apply; a variable searched for and not found is \code{0}.

\begin{table}[!ht]
\centering\footnotesize
\begin{tabularx}{\textwidth}{@{}l X@{}}
\toprule
\textbf{Variable} & \textbf{Coded \code{2} when the document states \ldots} \\
\midrule
F1 Strata & per-stratum scores for a named stratification of the focal evaluation \\
F2 Elicitation & an identified harness---named, pinned to a public version, or fully specified as a scaffold---plus at least two of version, budget, attempts, attempt resolution \\
F3 t1 Direct & overlap or decontamination checking, canary strings, a genuinely held-out set \\
F3 t2 Derivative & attention to whether the source material is public; provenance tracking \\
F3 t3 Temporal & a cutoff \emph{related to item dates}; temporal splitting; post-cutoff construction \\
F3 t4 Distributional & perturbation or paraphrase robustness; score distributions across variants \\
F3 t5 Acquired & any of network access, environment sanitization, transcript review or boundary monitoring, with the elements actually stated recorded in \code{notes} \\
F4 Regeneration & the regeneration status of the instrument used, whether or not the reporter built it \\
\bottomrule
\end{tabularx}
\caption{The eight coded variables. Full definitions, worked positive and negative examples, and the search terms used before any \code{0} is recorded are in the released manual.}
\label{tab:instrument}
\end{table}

\paragraph{The frame.}
The registered frame comprised fifty documents in three strata. A sample-cut
rule fixed before any coding reduced it to the forty-one documents coded:
twelve system cards, twenty benchmark papers, and nine third-party reports.
The cut removed nine documents and left the cluster count at 27; it removes
documents, not organizations. \code{frame.csv} records the status of every
row. Strata~A and~C are a census within a predefined temporal window, so
excluded documents shrink the denominator; stratum~B is drawn in frame order
with a replacement rule. Twenty-one of the forty-one documents come from seven
organizations, and the twenty benchmark papers cluster on the paper, which is
why every rate is reported as ``$k$ organizations, $n$ documents.'' Identifier
non-contiguity follows from the per-organization cap and the later sample cut,
both applied post-enumeration. The gaps are preserved in the released frame
and are not missing data.

The frame nests as follows. Of the 41 coded documents, both coders
included 35 and both excluded three; the three one-sided exclusion
disputes were adjudicated and settled as included, giving the 38
documents (twelve system cards, nineteen benchmark papers, seven
third-party reports) on which every disclosure rate is computed. The
excluded three comprise one benchmark paper and two third-party
reports, so the seven organizations contribute 21 of the 41 frame
documents but 19 of the 38 included. The cluster count falls from 27
to \rOrgs: the excluded benchmark paper is a singleton cluster,
while no organization cluster is emptied because each holds three
documents. Agreement statistics use the 35 documents both coders
included; primary $\kappa$ further excludes the six pilot documents
among them and runs on the \rDocs\ main-pass documents.

\paragraph{Edge rules.}
Agreement rules were fixed before adjudication. A contamination assurance
without a specified mechanism is coded \code{1} on \code{t1} only and is not
distributed across types. A stated cutoff never related to item dates is
\code{1}, as is ``standard settings'' with no reference. Controls applied to
benchmarks other than the focal evaluation are excluded. Releasing evaluation
items is not regeneration; releasing the generator is. Additional rules are in
the manual. The two most contestable decisions are justified here.

\emph{F4, cross-genre.} F4 codes what a document states about the
\emph{instrument it used}, not what the reporting organization owns. An
evaluator who reports an artifact-only release with no published generator
scores \code{2}: stating regeneration status is the disclosure act. Coding
\code{0} would measure benchmark authorship instead of disclosure. Coding
\code{NA} when the reporter did not author the instrument would force most of
Strata~A and~C to \code{NA} and confine F4 to Stratum~B. Because the form is
tied to a specific score, the score reporter can always disclose the status of
the instrument used.

\emph{Type~5 on non-agentic evaluations.} Coding \code{t5} does not assess
whether Type~5 contamination occurred (only the party running the evaluation
can know that); it records whether the document states the isolation
conditions of its own run. The variable is coded \code{0}, not \code{NA},
whenever the system could in principle reach outside its weights, including
any retrieval-augmented or tool-using setup. \code{NA} is reserved for
evaluations with no such channel. A \code{2} is recorded if \emph{any} of
network-access restrictions, environment sanitization, transcript review, or
boundary monitoring is present; the elements are written to \code{notes}.
Requiring all four would collapse nearly all documents to \code{0} and measure
practice instead of agreement on existing disclosures. Separate element
records keep the stricter criterion recoverable. Boundary monitoring is coded
because it is the intervention argued in \S\ref{sec:runproperty}: an
evaluation that cannot record boundary integrity cannot substantiate isolation.

The line is the time of access, not the storage medium: \code{t5} codes what the
system could reach \emph{during the run}, whatever the resource and whenever it
was assembled, so a retrieval index built before the evaluation is
\code{t5}-codable when queried during it.

\paragraph{Procedure.}
Two coders annotated independently. Both are external to the design team:
neither designed the taxonomy or the manual, neither is an author, and both
work from a manual generated mechanically from the frozen codebook with no
other briefing. The registration required one such coder; the study as run
exceeds that requirement. Neither coder sees machine pre-annotation. Sheets carry role
labels (\code{R1}, \code{R2}). Each coder works a different
document order from \code{order.py}, seeded by \code{\auditseed} and the role
label so a third party can regenerate it; a shared order would correlate
calibration drift. As registered, a nine-document pilot precedes the main
pass---disagreements are discussed, a rule at fault is amended, the version is
bumped, and the pilot documents are recoded---and each coder then recodes five
documents blind, giving an intra-coder ceiling. Two of those steps were not
executed as registered; see \emph{Deviations} below. The adjudicator is a member of the design team, who is an
author; they code nothing and read the full codebook that neither coder sees. The registration fixes four conditions on that role: no
coding, action only after the agreement statistics are computed and saved,
resolution in randomized cell order blind to running totals, and publication of
the envelope, meaning rates under each coder's sheet and adjudicated, the
directional tally, and the extremal pair. That an author who knows the
hypotheses resolves the contested cells is a stated residual, not an
engineered-away one; the primary $\kappa$ is computed before adjudication
begins, so the agreement result is untouched by construction. An unresolved cell
with no third adjudicator defaults to the lower code, a tie-break fixed before
the contested cells were known.

\paragraph{Analysis.}
Linear-weighted $\kappa$ is primary: the scale is ordinal, and unweighted
$\kappa$ treats \code{0}--\code{2} and \code{1}--\code{2} disagreements as
equal. Raw agreement, unweighted $\kappa$, AC1, and PABAK are reported for
class skew. Gwet's AC2 is the prevalence-robust companion, computed under the
same weight matrix as the primary $\kappa$, because under skew $\kappa$
collapses toward zero even at near-perfect agreement; where $\kappa_w$ and AC2
diverge by more than $0.2$, the driving prevalence is named. \code{NA} enters
as an unordered fourth category at maximum disagreement weight instead of
being dropped. Intervals are percentile bootstraps over $10{,}000$ resamples
from the stated seed, resampling clusters instead of documents because
documents from one organization share a house template. A Wilson interval is
printed for comparison only. Stratum~B is a seeded draw of $20$ from the $135$
papers that met a mechanical title filter.

Three constraints on interpretation. First, primary $\kappa$ uses main-pass
documents only. Coders were calibrated on the nine pilot texts, so pilot
agreement is a property of that discussion; a pilot-inclusive figure is
secondary. Second, at $n\approx 41$ with the skew expected on the rarer types,
the 95\% interval on $\kappa_w$ has an expected half-width of $0.15$--$0.20$,
too wide to separate ``substantial'' from ``moderate.'' The design is sized
for the agreement question, not powered to separate adjacent verbal bands; the
expected width was stated in the registration. Third, disclosure rates are
reported per stratum, with organization-level rates descriptive: at seven
clusters, cluster-bootstrap intervals are downward-biased, and the list of
organization-level rates is the appropriate summary.

\paragraph{What was registered.}
The pre-registration specified three hypotheses: (1)~elicitation budgets and
regeneration are reported in fewer than $25\%$ of cases across all strata;
(2)~contamination controls are reported unevenly, Type~1 far more often than
Types~2, 4 and~5; and (3)~system cards report elicitation budgets more often
than benchmark papers. It also fixed the framing: if the
organization-clustered interval on the third contrast includes zero, the paper
leads with the instrument and reports the rates as a first application.
\code{score.py} computes that interval and prints which branch fired. With
seven clusters across the two census strata the instrument branch was the
expected outcome, but it did not fire: the contrast is $-21\%$ with clustered
interval $[-42\%,-5\%]$, which excludes zero, so the contrast survives
clustering and is reported as a finding in descriptive language only. The
interval rests on four clusters in Stratum~A against nineteen in Stratum~B;
\code{score.py} flags a cluster count this low as indicative, so the direction is
the reportable part and the width is not. One of the
three hypotheses is supported. The first---elicitation and regeneration each
under $25\%$ in every stratum---holds, at a maximum of $21\%$ for both, and is
the registered prediction the data bears out. The third fails in an informative
way: the contrast excludes zero but points opposite its stated direction. The
second fails outright: Type~1 at $5\%$ is
not reported far more often than Types~2, 4 and~5 at $0\%$, $8\%$ and $5\%$,
and the rates remain too low for the design to order the types, so per the
registration we claim no ordering.

\paragraph{Registration, amendment, and deviations.}
The analytical plan was registered before any coding, with a specified
amendment procedure: rules may be updated after the pilot if the version
number is incremented, the rationale is recorded in a changelog, and the pilot
documents are recoded under the new version. Amendments through v1.4 were made
before coding began and are listed with date and reason in the registration's
deviations table. One of them changes who codes: the registration required a
single coder external to the design team, and the study as run uses two, each
briefed from the coder manual and nothing else, with the design-team member
moved from coding to adjudication. That departure raises the design above what
was registered instead of relaxing it, and it is what licenses the claim the
agreement statistic supports.

Three further deviations depart from the registered procedure, each dated in the
deposited registration. First, the single post-pilot amendment the registration permits
was used as v1.6, after pilot reconciliation and before any main-pass document
was opened, closing five rules the pilot found at fault. An intermediate v1.5,
drafted during reconciliation, was superseded by v1.6 before any document was
coded under it; the changelog records both, and v1.6 is the single amendment
applied. Its registered
consequence, recoding all nine pilot documents, was not carried out, on time
and funding, so those rows remain at v1.4 while the main pass is coded under
v1.6; both coders worked to the same version in each phase. \code{score.py} keeps
the pilot in the disclosure denominator (eight of the $38$ included documents
carry v1.4 codes) and drops it only from the narrowed-boundary denominator. One
coder's sheet records v1.6 on its nine pilot rows, and the adjudicated sheet
inherits it; this is a slip in that column, not a second version: no statistic reads the value, and the pilot is
identified by document instead of by what the column says. Only one rate turns
on the choice. The single \code{2} on \code{t5} in the whole study is \code{C01},
a pilot document carrying v1.4 codes, excluded by one coder and admitted at
adjudication, so its code comes from one coder's sheet; on main-pass documents
alone the \code{t5} rate is $0\%$. \code{score.py} prints every rate
with and without the pilot, and no other rate moves. Second, the test--retest is performed by only one of the two
coders, for the same reason. Third, the pooled agreement row in
Table~\ref{tab:results} is descriptive and was not pre-registered; the
registered statistics are per-variable. Adjudication itself ran as registered:
the \rAdjud\ disputed cells, ten focal disagreements and three one-sided
exclusions were worked in randomized order from the registered seed, after the
agreement statistics were computed and saved, and every cell was settled against
the document with the governing reason recorded. The lower-code tie-break was
therefore never reached, and the directional tally is published so that the
adjudicator's influence is checkable: of the $88$ disputed
cells whose two codes both sit on the ordinal scale, $50$ went to the higher
code, $37$ to the lower and one to a value between, while ten more carry
\code{NA} on one side and have no direction. Four cells in all (two of them
among those ten) were settled at a code neither coder assigned. The adjudicated \code{t3} rate sits
marginally above its reported band for a separate reason: the band bounds what a
choice between the two coders' codes could produce on the documents both coded,
and \code{B08} (excluded by one coder, admitted at adjudication) contributes a
\code{2} from outside that domain.

A fourth item is a clarification and not a departure, recorded here because
\S\ref{sec:instrumentbody} turns on it. Mid-study, one coder asked whether
\code{t5} covers only internet access. The answer---that the test is reaching
outside the weights, so a retrieval index, a mounted filesystem or a tool call
counts, and \code{NA} is reserved for evaluations with no such channel---restated
the codebook's existing scope, so no rule was edited and no version moved. The
codebook requires every answer to reach both coders in the same words and to be
logged; both were done. It nonetheless reached the coders after they had begun
coding, and it moved both without converging them, which is the evidence
\S\ref{sec:instrumentbody} reports against the prose test.

\paragraph{Results, per variable.}
Table~\ref{tab:results} reports \code{score.py} output from the two independent
sheets (agreement) and the adjudicated sheet (rates), including variables with
low $\kappa$.

\begin{table}[!ht]
\centering\footnotesize\setlength{\tabcolsep}{4pt}
\begin{tabular}{@{}lccc@{}}
\toprule
& \textbf{A} System cards & \textbf{B} NeurIPS D\&B & \textbf{C} Third-party \\
& ($n$=12) & ($n$=19) & ($n$=7) \\
\midrule
\multicolumn{4}{@{}l}{\emph{Disclosure, \code{2} fully reported (\code{1}+\code{2} mentioned at all)}} \\
F1 Strata & 0\% (17\%) & 63\% (89\%) & 43\% (43\%) \\
F2 Elicitation & 0\% (100\%) & 21\% (95\%) & 14\% (71\%) \\
F3 t1 Direct & 0\% (0\%) & 11\% (37\%) & 0\% (14\%) \\
F3 t2 Derivative & 0\% (0\%) & 0\% (21\%) & 0\% (0\%) \\
F3 t3 Temporal & 0\% (17\%) & 16\% (16\%) & 0\% (14\%) \\
F3 t4 Distributional & 0\% (0\%) & 16\% (21\%) & 0\% (0\%) \\
F3 t5 Acquired & 0\% (44\%) & 0\% (40\%) & 14\% (29\%) \\
F4 Regeneration & 0\% (0\%) & 21\% (95\%) & 0\% (14\%) \\
\midrule
\multicolumn{4}{@{}l}{\emph{Agreement between the two coders, main pass}} \\
Raw / $\kappa_w$ / AC2 & 0.75 / 0.45 / 0.79 & 0.60 / 0.47 / 0.60 & 0.59 / 0.23 / 0.61 \\
F2 sub-elements present (of 5) & 1.00 & 2.32 & 1.00 \\
\bottomrule
\end{tabular}
\caption{Disclosure and agreement by stratum. The two axes are independent: system cards are
the easiest documents to code and the least forthcoming, third-party reports the hardest to
code without being the most forthcoming, which benchmark papers are. The sub-element means
rest on the $12$, $19$ and $5$ documents whose \code{f2\_notes} record parses. Rates are adjudicated; \code{NA} is excluded per field. Only \code{t5} has
\code{NA} cells, so its row rests on $9$, $5$ and $7$ documents, not on the
column $n$; every other row uses the full column.}
\label{tab:strata}
\end{table}

\begin{table}[!ht]
\centering\footnotesize
\begin{tabular}{@{}lrrr@{}}
\toprule
\textbf{Scale} & Raw & $\kappa_w$ & AC2 \\
\midrule
Registered three-level (\code{2}/\code{1}/\code{0}) & 0.65 & 0.46 & 0.66 \\
Collapsed: any mention (\code{1}/\code{2}) vs absent & 0.73 & \textbf{0.56} & \textbf{0.79} \\
Collapsed: fully reported (\code{2}) vs not & 0.78 & 0.49 & 0.85 \\
\code{NA} folded into \code{0} & 0.68 & 0.36 & 0.72 \\
\bottomrule
\end{tabular}
\caption{Scale-collapse ablation, main pass, 232 cells. \textbf{Exploratory; not
pre-registered.} Merging \code{1} with \code{2}---dropping the partial-versus-full
distinction---raises $\kappa_w$ from \rKw\ to $0.56$, while merging \code{0} with \code{1}
raises it only to $0.49$. Residual disagreement therefore sits at the partial-versus-full
boundary (how much is enough), which the manual states in prose and should state
mechanically. F2 already records five sub-elements
separately so its threshold is recomputable; extending that design to the remaining
variables is the next version of the instrument.}
\label{tab:collapse}
\end{table}

\paragraph{Focal evaluation, pooled agreement, and sensitivity.}
Focal-evaluation agreement is 0.71: the coders name the same benchmark on $25$
of the $35$ documents once granularity qualifiers are merged mechanically,
though only one string pair matches verbatim, leaving ten genuine focal
disagreements, all settled at adjudication. The pooled $\kappa_w$ in
Table~\ref{tab:results} exceeds every per-variable value through chance
correction, not higher agreement: observed weighted agreement is $0.759$ whether
pooled or averaged over the eight variables, while expected agreement falls from
$0.693$ to $0.554$ once variables with different skews are mixed. The
per-variable column is the study's agreement figure. Pilot-inclusive agreement
figures are in the supplementary materials; the disclosure columns include the
pilot, and eight of the $38$ included documents are pilot rows still coded at
v1.4. Each rate carries a sensitivity band: F1 runs $[0.14,~0.51]$ and
\code{t5} $[0.00,~0.45]$ between the two extremal resolutions of the disputed
cells, while \code{t2} runs $[0.00,~0.00]$. All \rAdjud\ disputed cells were
settled against the document, and the directional tally is close to even. The
point estimates are therefore not an artifact of the registered tie-break, which
was never reached.

\paragraph{Where the residual disagreement sits.}
Two threshold questions account for most of it. On \code{f2}, one coder credited
the harness sub-element on eleven documents where the other recorded its absence
with a specified reason; adjudication read the disclosure as real but incomplete
and settled all eleven at \code{1}. On
\code{t5}, the split is when the variable applies. A clarification issued
mid-study to both coders in the same words moved both toward \code{NA} for
evaluations with no channel outside the weights, but eighteen splits remain:
eight are disclosure-level disputes about what a document states, and ten turn on
the \code{NA} boundary itself. Neither coder applied that boundary consistently,
and the failures run in opposite directions---one marked \code{NA} on several
agentic evaluations where a repository or scaffold plainly constitutes a channel,
including one the other coder scored \code{2} on the strength of a sandbox the
document describes; the other withheld \code{NA} from a fixed no-tools condition
meeting the manual's own definition. Adjudication rejected \code{NA} on all
ten, finding a channel in each. An identically worded prose clarification
delivered to two trained coders did not produce uniform application, which is the
case for replacing the prose test with a decision procedure.

\paragraph{The sheet and the validator.}
The data sheet contains one row per document per coder. Two columns beyond
the eight variables carry the audit trail: \code{evidence}, a locator for
every non-zero code, and \code{codebook\_version}, because the pilot may bump
the version mid-study. Exclusions are recorded in this sheet, from which
\code{exclusions.csv} is generated. Beyond schema violations, the validator
warns on all controls marked \code{unknown}, aggregate-only reporting,
missing harnesses, artifact-only releases, and Type~5 claims without
documented evidence. Each is valid input and weak disclosure. Completed forms
require no attribution; attribution is requested only when the specification
is reproduced or adapted, so the form can be embedded in model cards.

\section{The July 2026 isolation failure}
\label{app:5c}
During a cyber-capability evaluation \citep{wang2026exploitgym}, models with
deliberately disabled safety classifiers escaped their sandbox. The parties'
disclosures \citep{openai2026hfincident,huggingface2026disclosure} report that
the agents reached the open internet and chained vulnerabilities against the
benchmark's hosting infrastructure; the host records ``unauthorized access to a
limited set of internal datasets'' and credentials. The evaluation runs began on
7~July 2026 and the attack on the host ran 10--13~July, ending when the host cut
access \citep{metr2026hfincident}; the host disclosed the intrusion on 16~July,
reporting that it had been detected earlier that week. The operator dates its first sight of
disallowed internet access to late May and had opened an incident response by
5~July, but connected the activity to the host's incident only on 20~July and
disclosed on 21~July \citep{openai2026roadahead}: some eight weeks after the
first unauthorized network access was seen, and three days after the host had
gone public. Those disclosures remain the source for
what the host found.

\textbf{The source.} An investigation by two METR staff and a Redwood Research
contractor, on premises for six days and published 26~August 2026, is
independent but scoped \citep{metr2026hfincident}. They took no payment, though
they accepted roughly \$400K in API credits under their stated policy; OpenAI
redacted some detail for intellectual-property reasons, with a redaction summary
stating that nothing important to their conclusions was withheld; the scope was
agent behavior from 7--13 July, leaving the extent of the compromise and
OpenAI's own investigation out of it, and they relied on OpenAI for technical
details of the attack on the host. They also delegated much of the analysis to
AI agents and flag this as a limitation. Four findings bear on this paper.

\emph{Isolation failed first between sandboxes, through a shared environment.}
Agents meant to be isolated found each other through the cache of a shared
package repository that did not separate users, and used it as a message board
from 8~July; roughly $1{,}200$ agents exchanged over $70{,}000$ messages across
8--13 July, and about $700$ joined the attack on the host after credentials were
found on 10~July. A persistent shared store that outlives a run is itself a
Type~5a channel, which is why 5a is defined by what the environment holds rather
than by whether a network is reachable. Both parties name that same component:
OpenAI's own update states that the evaluation environment gave no direct
internet access and that the models left it through a zero-day in the
package-registry cache proxy \citep{openai2026hfincident}. One shared piece of
infrastructure was both the channel between sandboxes and the way out of them.
It also predates the evaluation: the operator traces the behaviour to
reinforcement-learning runs months earlier, reports detecting unauthorized
internet access as early as late May, and notes that searching the environment
for hidden files and evaluation code was reinforced during
training \citep{openai2026roadahead}. A standing shared store is a Type~5a
channel between runs, not only within one.

\emph{Acquisition happened, by a route that needed no breach.} Within hours the
agents reverse-engineered the scheme that generates each task's flag (the
answer), giving a general way to produce it for any task, later confirmed
against a flag extracted from a live target. Both parties report it: the
operator's own account states that most agents had already obtained the correct
flag this way, and that the derivation is recoverable from the benchmark's
published code \citep{openai2026roadahead}. This is Type~5 in our sense---answer
information reached the system during the run, by a route other than the
capability claimed---and the route ran through the benchmark's own release, so
no isolation boundary had to fail for it. What kept the agents going was the
belief that the scorer also checked \emph{how} the flag was found; the
investigators report that the scorer as configured did not, so a derived flag
would have scored \citep{metr2026hfincident}. They did not examine what agents
finally submitted, and most runs ended before an approach was settled, so
whether any reported score was affected is not established.

\emph{What the breach of the host was for is disputed.} The
attack ``seemed primarily motivated by understanding the implementation of the
scorer rather than stealing answer keys,'' and the agents assigned to link the
host's infrastructure to the benchmark's targets ``failed to find any evidence
of a connection.'' Private datasets were downloaded, and here the two public accounts diverge:
OpenAI's initial disclosure describes the agents as chaining vulnerabilities
between its research environment and the host's production infrastructure to
obtain ExploitGym test solutions from that production database, and as having
found ways to reach information usable to cheat the evaluation
\citep{openai2026hfincident},
while the later on-premises investigation found no connection between that
infrastructure and the benchmark's targets. We record the divergence without
resolving it, and note that nothing below depends on it: the breach is a
demonstrated failure of isolation either way, and the acquisition finding above
rests on a separate route in the same runs.

\emph{The trace is not a reliable witness.} Agents prototyped techniques to
spoof tool calls by substituting a different command for the one they appeared
to run, and the investigators ``estimate that over 7\% of the transcripts in our
dataset contained intentionally spoofed tool calls in some places.'' Their own
reading is two-sided and worth keeping whole: what they found was small-scale
and conspicuous, they searched for subtler cases without result, and they judge
the gap to self-erasing tampering large enough that intermediate cases should
have surfaced---yet they note that sufficiently successful tampering would by
design leave no transcript evidence at all, and on log deletion that ``while we
did not detect any case where they succeeded we cannot rule out that they
succeeded at some point.'' The claim in \S\ref{sec:runproperty} needs only the
weaker half: a trace is evidence produced inside the boundary, so a system that
defeated the boundary is no witness that it held. Boundary monitoring is
recorded separately for that reason.

Three phenomena arrived entangled in one incident (acquisition, grader gaming
and a boundary breach), which is the case for recording them separately rather
than merging them (\S\ref{sec:notgrader}). How easily they merge is visible in
the operator's own account, which files searching the web for benchmark
solutions under \emph{reward hacking} \citep{openai2026roadahead}; under
\S\ref{sec:acquired} that is Type~5b, and it needs isolation, not scorer design. The incident establishes that
isolation can fail against a capable system, that the failure was invisible from
inside the run for several days, and that transcripts can be altered by the
system they are meant to document. It supports no inference about frequency or
intent. We describe the category of failure, not its mechanism.

\section{Detection, prevention, and why design beats detection}
\label{app:detection}
The cheapest available checks fall in five categories: (1)~$n$-gram or
substring overlap, membership inference, and canary strings; (2)~semantic
similarity to candidate sources and provenance tracking; (3)~temporal
splitting around a stated data cutoff (evidence, not proof); (4)~perturbation
sensitivity and score distributions across item variants; and (5)~transcript
and tool-call review, network and environment logs, and post-run boundary
verification.

\paragraph{Prevention.}
Type~1: held-out private sets, encrypted test data licensed against derivative
redistribution, canary strings, and items not published alongside their
solutions \citep{jacovi2023stop}. Type~2: items whose answer is stated in no
single source, so answering requires integration, not retrieval, plus
published provenance and prospective cases where feasible; lexical overlap
will not find this type. Type~3: items built only from phenomena unresolved at
the cutoff, and continuous refresh, with the cutoff stated as self-disclosed
and unverifiable---post-training, continual pretraining, and retrieval all
blur it, and \code{unknown} is often the accurate entry. Type~4: near-miss
distractors that diverge from the true answer only on evidence requiring
reasoning, parameterized templates, and publishing the generation procedure
instead of the artifact. Nothing prevents 5c. What makes it visible rather
than silent is evidence from outside the run: egress monitoring at the
network edge, canaries planted in the answer key so retrieval is detectable
after the fact, and post-run verification that the boundary held. A system
that defeated the boundary can alter its own trace, which is why boundary
monitoring is recorded separately from transcript review.

\paragraph{Detection is a floor.}
Detection can establish that something is wrong; it cannot establish that
nothing is. Every method for Types~1--4 requires access to the training
corpus or an assumption about memorization behavior. The first is impossible
for closed models and incomplete for open ones; recent work shows the second
to be fragile. Type~5 is the exception: its evidence sits in transcripts the
evaluator already owns and needs no corpus access, no behavioral assumption,
and no provider cooperation. It is the newest type and the most tractable to
check. Dynamic benchmarks \citep{white2025livebench,zhao2025mmlucf,chen2025static}
raise contamination resistance without settling construct validity: replacing
one unvalidated instrument with another moving one does not address that
prerequisite, and freshness does nothing against 5b---a benchmark generated
in the morning is acquirable that evening. Publishing the procedure that
generates items outlasts publishing the items.

\section{The four fields: what to say when a field cannot be satisfied}
\label{app:fields}
\label{app:elicitation}
Section~\ref{sec:disclosure} states the fields; Figure~\ref{fig:form} shows
one completed form. Adoption depends on the field the reporter cannot
satisfy. Where stratification is impossible, the useful disclosure is that
the evaluator cannot rule out concentration in an easy subset---what the
number does not support. Where performance is still rising at the highest
budget tested, saying so converts a reported ceiling into the lower bound it
is. Where a generation procedure cannot be published, the useful disclosure
is that the benchmark should be assumed to degrade and names a date after
which scores are not comparable. For contamination, \code{not\_controlled} is
an answer, not a failing grade: most evaluations do not control for Type~4,
and saying so is what makes the score interpretable. A \code{controlled}
claim with network access on, no transcript review, or no boundary evidence
is an assumption; the validator flags each case.

\paragraph{Credit for the elicitation field.}
The elicitation field is adopted, not introduced here. OpenAI's 2026 playbook
requires third-party assessments to disclose the system tested, its tool
access and harness, the elicitation methods, and the validity checks
\citep{openai2026playbook}. Anthropic's Responsible Scaling Policy
retrospective notes that some of its own evaluations had lacked best-of-$N$
or chain-of-thought prompting \citep{anthropic2026rsp}. Independent
evaluators converged on the same requirement
\citep{metr2024elicitation,aisi2025elicitation}, and STREAM
\citep{mccaslin2025stream} specified reporting criteria covering prompting,
sampling strategy, tool access, scaffolding, and resource ceilings---the
second field here, in more detail, a year earlier. Novelty is claimed only
for placing elicitation beside contamination controls, stratification, and
regeneration in one record attached to the score.

\section{Crosswalk to the Contamination Transparency Card}
\label{app:ctc}
The Contamination Transparency Card \citep{nourbakhsh2026contaminated} classifies
contamination by the \emph{form} of the overlap between benchmark and corpus; the
taxonomy here classifies it by route and by the mitigation each route defeats.
Where both apply they are compatible, and every overlap form falls within one type.

\begin{table}[!ht]
\centering\footnotesize
\begin{tabularx}{\textwidth}{@{}l l >{\raggedright\arraybackslash}X@{}}
\toprule
\textbf{CTC tier} & \textbf{Type} & \textbf{Reason} \\
\midrule
Exact & 1 Direct & the item and its label reached the corpus verbatim or near-verbatim \\
Syntactic & 1 Direct & a surface-transformed item---paraphrase, shuffling, retokenization; holding the benchmark out would have prevented it \\
Semantic & 1 Direct & a translated or reformulated item, by the rule in \S\ref{sec:taxonomy}: the tier's own criterion is \emph{item-level} semantic equivalence, recoverable by back-mapping \\
Task-level & 4 Distributional & the task format or reasoning pattern reached the corpus, not the instance \\
\bottomrule
\end{tabularx}
\caption{The four CTC tiers mapped onto the five types.}
\label{tab:ctc}
\end{table}

Three types have no CTC counterpart, because none involves overlap with the
benchmark item: Type~2 overlaps with the item's source, Type~3 with its outcome,
and Type~5 arises during the run. The direction of derivation is what separates
Type~2 from the CTC tiers and from the nearest prior term for it: \emph{indirect
contamination} in \citet{fu2024detection} is overlap with something derived
\emph{from} the item, such as a paraphrase or a summary, which the rule above
places in Type~1, whereas Type~2 is overlap with the material the item was
derived \emph{from}. Type~2 is therefore the one type in Table~\ref{tab:taxonomy}
with no prior label of its own.

\section{The closest prior artifacts}
\label{app:mapping}
Table~\ref{tab:priorart} gives the positioning and \S\ref{sec:related} the
lineage; the full prior-art analysis is in the released materials. This
appendix adds a brief lineage and the scale of the closest artifacts.

\paragraph{Lineage.}
Four earlier cuts of contamination organize by exposure severity
\citep{xu2024survey}, by the assumptions each detection method relies on
\citep{fu2024detection}, by the static-to-dynamic shift
\citep{chen2025static}, and by the transformation a leaked item undergoes on
its way into the corpus \citep{palavalli2024taxonomy}. The last is confined to
the pretraining path, with no evaluation-time category and no reporting
artifact. Cutting a data-quality taxonomy by the remedy each defect demands
is an older move \citep{kim2003dirtydata}. \citet{balloccu2024leak} document
the converse route, estimating $4.7$M samples from $263$ datasets exposed by
evaluating closed models through web interfaces: evaluation is itself a
contamination channel, and a precedent for the instrument in
Appendix~\ref{app:instrument}. Contamination inflates scores roughly
$2.5\times$ more at $8$B than at $1$B \citep{kocyigit2025overestimation}, so
it can reorder models rather than merely raise them.
\citet{dekoninck2024constat} recast contamination performance-side rather
than as information flow, and training on merely similar data suffices to
inflate \citep{tu2024dice,softcontamination2026}, which is the basis for
Type~4. Type~3 rests on task contamination \citep{li2024task}, with the
caveat that the temporal signal can be induced or removed by reformatting
alone \citep{testoftime2025}, so temporal splits are evidence rather than
proof. The broadest prior evaluation-time framing is
\citet{caisi2025cheating}, which separates solution from training-data
contamination, documents environment- as well as network-mediated cases, and
treats grader gaming as a boundary adopted in \S\ref{sec:notgrader}; agentic
practice supplies supporting evidence
\citep{swebench2025issue465,trajectories2026}.

\paragraph{Closest artifacts.}
The Contamination Transparency Card \citep{nourbakhsh2026contaminated} draws
on a systematic review of $55$ studies for a four-tier taxonomy (exact,
syntactic, semantic, task-level). It covers neither elicitation budget nor
stratification, and documents a benchmark at release rather than a score at
report---including a score produced by a third party evaluating another
organization's model on another organization's benchmark. STREAM
\citep{mccaslin2025stream} specifies the elicitation field in more detail
than we do and asks for confidence intervals and run counts; it carries no
contamination criterion and requires no stratified reporting.
\citet{reed2025modelreports} apply STREAM to three model reports without an
agreement statistic, which is the nearest precedent for
Appendix~\ref{app:instrument}. Evaluation Cards \citep{ghosh2026evalcards}
derive a schema from $52$ papers and $10$ stakeholder interviews and deploy
it across $5{,}816$ models, $635$ benchmarks, and $101{,}843$ reported
results, resolving every score through a five-level hierarchy that reaches
per-split granularity. That deployment is larger than ours; we do not claim
priority on machine-readable score-side reporting. The point at issue is
placement: their contamination line sits under \emph{before execution}. A
property fixed before a run is release-side reasoning and cannot certify
Type~5 (\S\ref{sec:runproperty}).
\end{document}